\documentclass[]{article}
\usepackage{emulateapj}
\usepackage{epsfig}
\usepackage{ifthen}

\submitted{Submitted to the Astrophysical Journal}

\def\gsim{\;\rlap{\lower 2.5pt
 \hbox{$\sim$}}\raise 1.5pt\hbox{$>$}\;}
\def\lsim{\;\rlap{\lower 2.5pt
   \hbox{$\sim$}}\raise 1.5pt\hbox{$<$}\;}
\def\msol{{\rm\,M_\odot}}

\def\kpc{{\rm\,kpc}}
\def\kms{\rm\,km\,s^{-1}}

\def\mic{{\,\mu{\rm m}}}

\def\kms{{\rm\,km\,s^{-1}}}

\def\spose#1{\hbox to 0pt{#1\hss}}
\def\lta{\mathrel{\spose{\lower 3pt\hbox{$\mathchar''218$}}
     \raise 2.0pt\hbox{$\mathchar''13C$}}}
\def\gta{\mathrel{\spose{\lower 3pt\hbox{$\mathchar''218$}}
     \raise 2.0pt\hbox{$\mathchar''13E$}}}

\def\zsol{{\,Z_\odot}}

\begin{document}
	
\title{Enrichment of the Intergalactic Medium by Radiation Pressure 
Driven Dust Efflux}

\author{Anthony Aguirre,\footnote{Institute for Advanced Study, School of Natural Sciences, Princeton NJ 08540}$^{,b}$
Lars Hernquist,\footnote{Department of Astronomy, Harvard University
60 Garden Street, Cambridge, MA 02138}
Neal Katz,\footnote{Department of Physics and Astronomy, 
University of Massachusetts, Amherst, MA 98105}
Jeffrey Gardner,\footnote{Department of Astronomy, University of Washington, 
Seattle, WA 98195}
\& 
David Weinberg\footnote{Department of Astronomy, Ohio State University, 
Columbus, OH 43210}}

\setcounter{footnote}{0}

\begin{abstract}

	The presence of metals in hot cluster gas and in Ly$\alpha$
	absorbers, as well as the mass-metallicity relation of
	observed galaxies, suggest that galaxies lose a significant
	fraction of their metals to the intergalactic medium (IGM).
	Theoretical studies of this process have concentrated on metal
	removal by dynamical processes or supernova-driven winds.
	Here, we investigate the enrichment of the IGM by the
	expulsion of dust grains from galaxies by radiation pressure.
	We use already completed cosmological simulations, to which we
	add dust assuming that most dust can reach the equilibrium
	point between radiation pressure and gravitational forces. We
	find that the expulsion of dust and its subsequent (partial)
	destruction in the IGM can plausibly account for the observed
	level of C and Si enrichment of the $z=3$ IGM.  At low-$z$,
	dust ejection and destruction could explain a substantial
	fraction of the metals in clusters, but it cannot account for
	all of the chemical species observed.  Dust expelled by
	radiation pressure could give clusters a visual opacity of up
	to $0.2-0.5\,$mag in their central regions even after
	destruction by the hot intracluster medium; this value is
	interestingly close to limits and claimed observations of
	cluster extinction. We also comment on the implications of our
	results for the opacity of the general IGM.  Finally, we
	suggest a possible `hybrid' scenario in which winds expel gas
	and dust into galaxy halos but radiation pressure distributes
	the dust uniformly through the IGM.

\end{abstract}
\keywords{cosmology: theory --- intergalactic medium --- galaxies: abundances
 --- dust: extinction}

\section{Introduction}
\label{sec-intro}

	Several independent sets of observations indicate that
galaxies must lose a substantial fraction of the metals they produce
during their lifetimes.  First, metal lines in hot X-ray emitting gas
in clusters and groups indicate that as much metal lies outside of
galaxies in these objects as inside them (e.g., Mushotsky et al. 1996;
Renzini 1997; Davis, Mulchaey \& Mushotsky 1999; Buote 2000).  Second,
quasar absorption line studies imply that the intergalactic medium
(IGM) at $z \la 3$ is enriched to metallicity $Z \ga 10^{-2.5}\zsol$
(e.g., Songaila \& Cowie 1996; Cowie \& Songaila 1998;
Ellison et al. 2000; Penton, Sticke \& Schull 2000).  Cosmological
simulations indicate that this seems to require at least $\sim10\%$ of
galactic metals to be ejected (Aguirre et al. 2001a,b).  Third, the
strong positive correlation between galaxies' masses and metallicities
(e.g., Zaritsky, Kennicutt \& Huchra 1994) is naturally explained
by the efficient escape of metals from low-mass galaxies (Dekel \&
Silk 1986; Lynden-Bell 1992).

	Most theoretical studies addressing this ubiquitous presence
of intergalactic metals have focused on the removal of metal enriched
gas from galaxies; the gas may be removed by ram-pressure stripping,
during dynamical encounters between galaxies, or as an outflow driven
by supernovae and stellar winds.  While dynamical removal undoubtedly
occurs at some level (especially in rich clusters), it is not clear
that it can account for the level of metallicity in the $z=3$ IGM or
the mass-metallicity (M-Z) relation of galaxies (Aguirre et al. 2001a;
but see Gnedin 1998).  Metal ejection by galactic winds can explain
the M-Z relation (winds escape low-mass galaxies more easily) and may
account for the observed level of IG enrichment (e.g., Cen \& Ostriker
1999; Aguirre et al. 2001b), but it is unclear whether they can do
this without overly disturbing the thermal or structural properties of
the high-$z$ IGM.

	A third metal removal mechanism, which has not
previously been treated in a cosmological context, is the ejection of
dust grains by radiation pressure.  As first pointed out by Pecker
(1972) and Chiao \& Wickramasinghe (1972), bright galaxies can 
exert a radiation pressure force on nearby grains that exceeds their
gravitational attraction, forcing the grains into the galaxies' halos
or beyond.  Subsequent studies involving realistic model galaxies have
confirmed this idea, showing also that gas drag is insufficient to
confine grains unless they start at small galactic scale-height (e.g.,
Ferrara et al. 1990; Shustov \& Vibe 1995; Davies et al. 1998;
Simonsen \& Hannestad 1999). 

	All of these studies support the idea that much of a galaxy's
dust may be ejected during its lifetime, so it is interesting to
assess the possible IG enrichment that would ensue.  Unlike winds,
enrichment by dust (partially destroyed in transit or by the IGM)
would not impact the thermal/structural properties of the IGM or
galaxies.  In this Letter, we assess the amount and distribution of
metals transferred to the IGM as dust driven by radiation pressure,
using two smoothed-particle hydrodynamics (SPH) simulations. The first
has $128^3$ dark matter particles and $128^3$ SPH particles in a
$(17\,{\rm Mpc})^3$ box, and ends at $z=3$.  The second, ending at
$z=0$, has $2\times 144^3$ particles in a $(77\,{\rm Mpc})^3$ box.
Both assume $\Omega_\Lambda=0.6$, $\Omega_b=0.047$, $\Omega_m = 0.4$,
$h=0.65$ and $\sigma_8=0.8$.  The simulations are described in more
detail in Aguirre et al. (2001a) and in Weinberg et al. (1999).
Section~\ref{sec-method} describes the method of adding metals and
dust to the already completed simulations. Section~\ref{sec-results}
gives results pertaining to the enrichment of the $z=3$ IGM and the
$z=0$ intracluster medium, in several representative models.  We
discuss these results and their implications in
\S~\ref{sec-discussion}.

\section{Method}
\label{sec-method}

The method by which we calculate IGM enrichment is discussed in
detail in Aguirre et al. (2001a).  Briefly, our method post-processes
a limited number of outputs from already completed SPH cosmological
simulations that include star formation.  We assume that each unit of
forming stellar mass instantaneously generates $y_*$ units of metal
mass.  We then deposit this metal mass in gas particles near the
forming star particle as follows:
\begin{enumerate}
\item{A fraction $(1-Y_{\rm ej})$ of the metal is distributed in the
nearest 32 gas particles, using the SPH smoothing kernel (see
Hernquist \& Katz 1989).  Half of the locally-distributed metal is
added in the form of dust, the other half as gaseous metal.}
\item{The remaining mass is tallied for a given galaxy,\footnote{By
`galaxy' we mean a group of bound particles found using the SKID
package, publicly available at
http://www-hpcc.astro.washington.edu/tools.} for which we also compute
the mean metallicity $\langle Z\rangle_{\rm gal}$ and the
UV-optical-NIR luminosity, using the models of Bruzual and Charlot\footnote
{The models are available via anonymous FTP from ftp.noao.edu.}
and a Scalo or Salpeter initial mass function (IMF).}
\item{Using $\langle Z\rangle_{\rm gal}$ we apply a dust correction to
the luminosity from Heckman et al. (1998; see Aguirre et al. 2001a),
normalized to give the observed ratio at $z=0$ in the cosmic
UV-optical-NIR and FIR backgrounds (which are also output by the
simulations).}
\item{We assume a grain size distribution (GSD) in mass $dm(a)/da$ and
opacity (per unit mass) law from Kim, Martin \& Hendry (1994) and Laor
\& Draine (1993), respectively, for either graphite or silicate
grains.}
\item{The fraction $Y_{\rm ej}$ of metal formed in a galaxy is
distributed as dust spherically about the center of star formation.  A
dust mass proportional to $dm(a)/da$ is placed in a shell where the
radiation pressure on a grain of radius $a$ balances the galaxy's
gravitation.}
\end{enumerate}

	The process is repeated for each galaxy at each time step. New
stars are formed with the metallicity (including dust) of the gas from
which they form.  Each gas particle has an accumulated mass of gaseous
metals and dust, and we track the GSD for each particle using a
9-point piecewise power law fit (see Aguirre et al. 2001a for
details).  The GSD is modified as the dust is converted to metals by

\vbox{ \centerline{ \epsfig{file=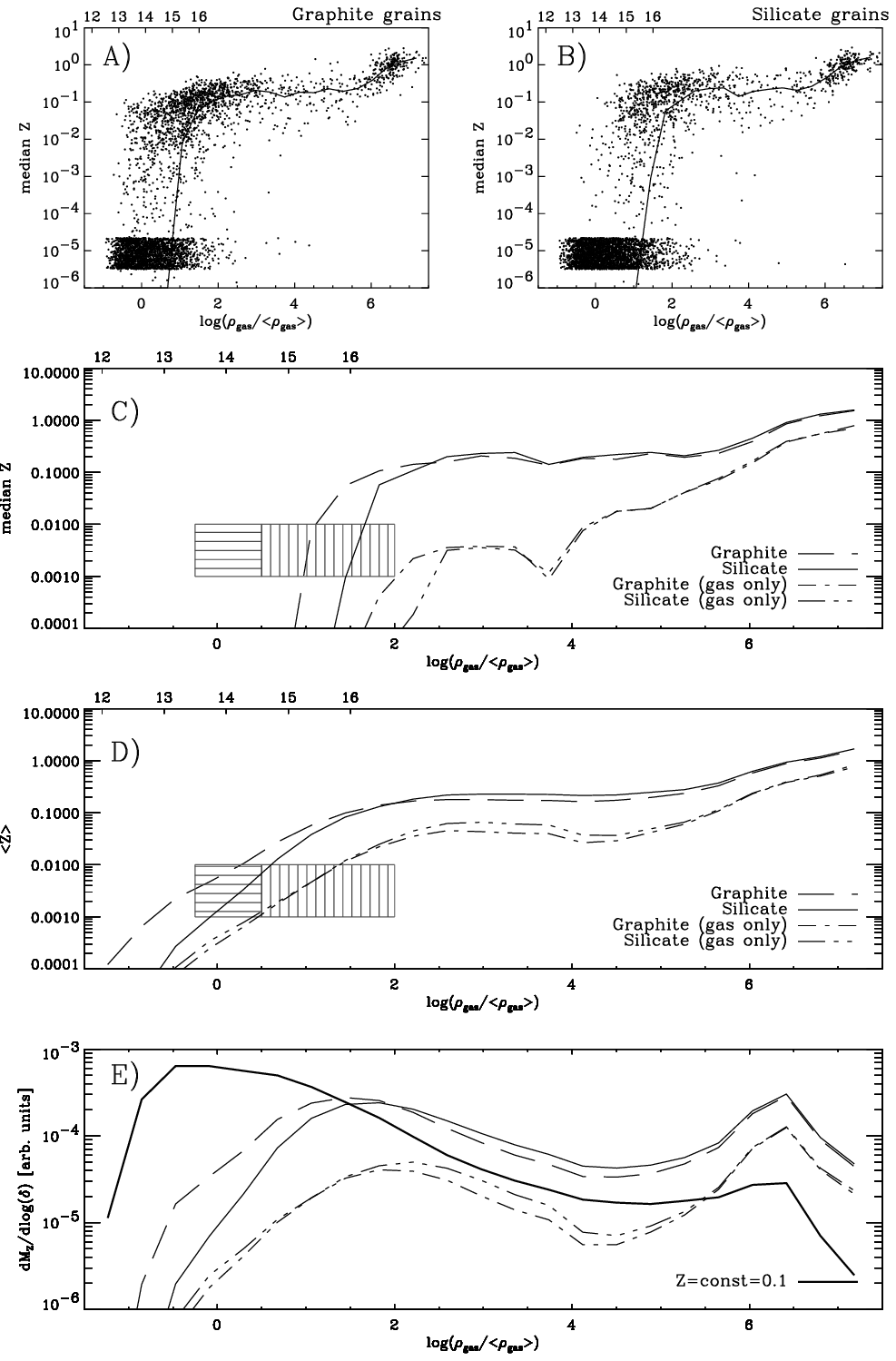,width=9.0truecm}}
\figcaption[]{ \footnotesize Enrichment of the IGM plotted in four
ways. {\bf Panel A:} Random subsample (1 in 500) of particle
metallicities for the fiducial model with graphite grains, versus
overdensity $\delta \equiv \rho_{\rm gas}/\langle\rho_{\rm
gas}\rangle$.  Top axis (here and in all panels) gives approximate
$\log N(H\,I)$, using the relation of Dav\'e et al. (1999).  The solid
line shows the median metallicity versus $\delta$. {\bf B:} As for
panel A, for silicate grains. {\bf C:} Median metallicities versus
$\delta$ for models with graphite and silicate grains, but for total
(dust+gas) metal content, and for gaseous metals only.  
The shaded box roughly indicates the 
metallicity of low-column density Ly$\alpha$ absorbers (Ellison et al. 2000). 
{\bf D:} As for panel C, but {\em mean}
metallicities are plotted. {\bf E:} As for panel C, but gives mean
metallicities times the fraction of baryons at a given $\delta$,
showing the contribution by components with different $\delta$ to the
cosmic metal density. The thick line shows the distribution assuming
constant metallicity (with the same total metal mass).
\label{fig-dustres}}}
\vspace*{0.5cm}

\noindent thermal sputtering by the IGM (using the yields of Jones et al. 1994),
or as new (unsputtered) dust is added to the particle.

\section{Results}
\label{sec-results}

	Our basic model assumes graphite grains, a 1:1 ratio between
the cosmic UV-optical-NIR and FIR backgrounds 
at $z=0$ (c.f. Madau \&
Pozzetti 2000), a Scalo IMF with cutoffs at 
$0.1\msol$ and $100\msol$,
$y_*=\zsol$, and $Y_{\rm ej}=0.5$.  The last assumption is maximal, as
only $\sim 1/2$ of a typical 

\vbox{ \centerline{
\epsfig{file=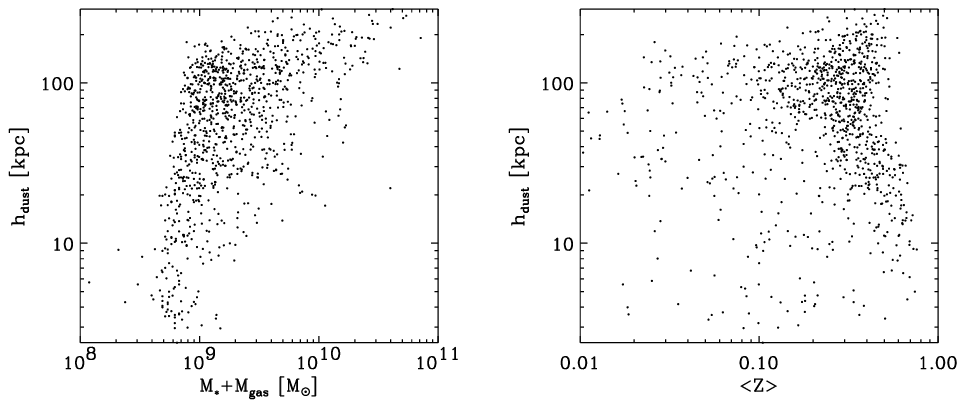,width=9.0truecm}} \figcaption[]{ \footnotesize
{\bf Left:} Maximal dust ejection radius $h_{\rm dust}$ vs. galaxy
mass for $z=3$. {\bf Right:} $h_{\rm dust}$ vs. mean metallicity. 
\label{fig-pqz3}}}
\vspace*{0.5cm}

\noindent galaxy's metals are in dust.  We also
give corresponding results for silicate grains.

	Figure~\ref{fig-dustres} shows the key results at $z=3$, using
the $128^3$ simulation.  
The top two panels give a sparse sampling of individual particle
metallicities, versus the gas overdensity $\delta$.  The stellar yield
$y_*$ is uncertain by perhaps a factor of two, and all of the curves
could be scaled vertically for a higher assumed value.
The metallicity at $\delta \la 10^4$ could
also be (roughly) scaled by $Y_{\rm ej}$ for lower assumed values.
The bar at the bottom of each panel shows the zero metallicity
particles and indicates that the distribution is rather inhomogeneous,
especially for silicate grains (Panel B).  This can also be seen by
comparing panels C and D, which show the median and mean metallicity
vs. $\delta$.  The latter shows that dust ejection can provide {\em
enough} metals to account for the Ly$\alpha$ observations (indicated
by the hatched rectangle), though the enrichment may not, in these
models, be uniform enough.  It is important to note, however, that
(assuming grains decouple from the galactic gas) our method always
{\em under}estimates the radius to which the grains can escape,
because they would inevitably reach the force balance radius with some
velocity and overshoot it.  Thus the distribution should probably be
more uniform than shown here. (The ejection radius should also be
limited by the average dust velocity $\bar v_d$ times the available
time, but introducing this limit does not change the fiducial model
results unless $\bar v_d \la 100\kms$, slow compared to velocities
seen in more detailed studies of dust ejection.)

Figure~\ref{fig-dustres} gives results for both the {\em total} metal
enrichment (dashed and solid lines), and for the gas-phase enrichment
(single- and triple-dot-dashed lines), where grains have been
converted to gas by thermal sputtering only.  Because destruction by
both thermal and nonthermal sputtering during grain ejection would
destroy more dust, true gas-phase abundances should lie above the
latter two curves (although if grains are destroyed {\em very}
efficiently at small radii they will not survive to pollute the
low-density regions).

The models with different dust corrections (e.g. changing the 1:1
ratio in cosmic backgrounds to 1:2 or 3:1) give differences in $z=3$
enrichment comparable to the 

\vbox{ \centerline{ \epsfig{file=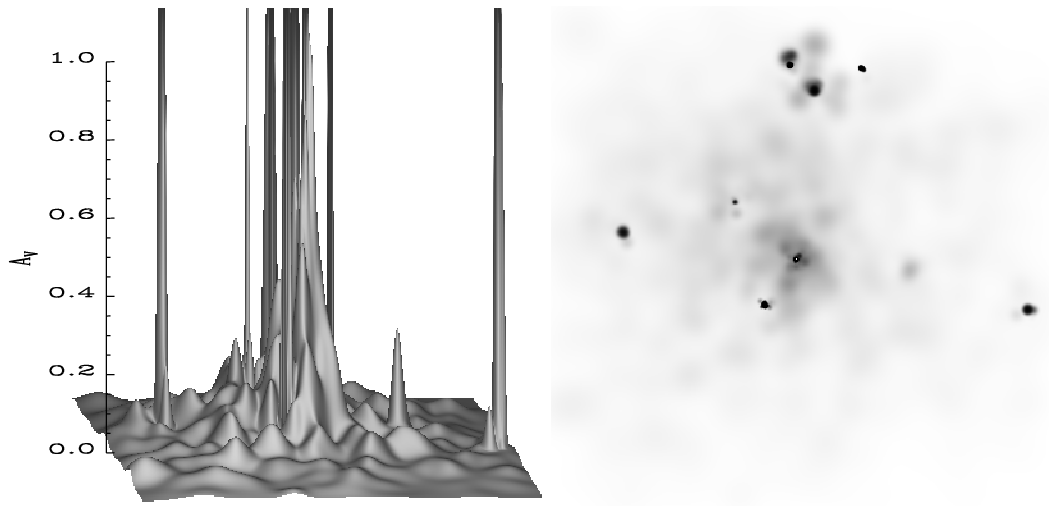,width=9.0truecm}}
\figcaption[]{ \footnotesize Dust extinction for a simulation
cluster at $z=0$ in the fiducial model, with graphite grains.  Both images are
1.2\,Mpc$\times$1.2\,Mpc, projected through a 1.2\,Mpc cube. The left
surface gives visual extinction, assuming $\kappa_V=4\times 10^4{\rm
\,cm^2\,g^{-1}}$.  The right panel simulates what a sheet of white
paper would look like through the dust of the cluster.
\label{fig-twoclust}}}
\vspace*{0.5cm}

\noindent differences between the silicate and
graphite models (which differ in dust opacity by a factor of a few).
Similar changes are induced by different assumed IMFs (see Aguirre
et al. 2001a).  

Quantities pertaining to the galaxies ejecting dust at $z=3$ are shown
in Fig.~\ref{fig-pqz3}.  The left panel, giving the maximal dust
ejection radius vs.\ the galaxy (baryon) mass, shows that ejection is
most effective from the larger galaxies.  This, and a correlation
between mass and metallicity (resulting from more efficient star
formation in larger galaxies), largely washes out the anti-correlation
between ejection radius and metallicity one would expect from the
metallicity-dependent dust correction (as shown in the right panel).
At low redshift, the balance reverses, and high mass galaxies eject
metals slightly less efficiently due to their large dust corrections.

	The $144^3$ simulation (which runs to $z=0$) allows us to
assess the enrichment of the low-$z$ IGM by dust ejection.  This
simulation only resolves galaxies of baryon mass $\ga 10^{10.7}\msol$,
but these galaxies dominate the observed $z=0$ mass function, and the
more efficient ejection of grains from large galaxies at high-$z$
(when small galaxies contribute relatively more mass), so we capture
the bulk of the metal enrichment.  In the fiducial model described
above, the ICM of the most massive groups/clusters is enriched to
$\approx 1/5\zsol$.

Dust is destroyed efficiently in the hot ICM, but enough remains that
some extinction can occur. Figure~\ref{fig-twoclust} shows the optical
depth through a simulation group/cluster of baryonic mass $2\times
10^{13}\msol$ in the fiducial graphite model, assuming a dust visual
opacity of $\kappa_V=4\times 10^4\,{\rm cm^2\,g^{-1}}$ (reasonable for
a more realistic silicate-graphite mixture). Except along paths
through galaxies, the cluster optical depth is typically $\la
0.2\,$mag; the central region has $A_V \sim 0.5$\,mag.  Poorer groups
show less opacity.  We note also that the same model predicts a
general `diffuse' extinction to $z=0.5$ of $\approx
0.05-0.1(\kappa_V/4\times 10^4\,{\rm cm^2\,g^{-1}})$\,mag, which is
comparable to the difference between Hubble diagrams for different
cosmological models at $z=0.5$ (Aguirre 1999) and could potentially be
important in observational cosmology.\footnote{The far-infrared
emission from such dust would not violate constraints from the
observed far-infrared or microwave backgrounds; see Aguirre \& Haiman
(1999).}

\section{Discussion and Implications}
\label{sec-discussion}

In \S~\ref{sec-intro} we argued that dust ejection is an interesting
alternative to dynamical or wind enrichment of the IGM because it may
be efficient, yet not disturb the IGM or galaxies in a way
incompatible with observations.  Our simulations, which produce
reasonable predictions for the masses, luminosities, and spatial
distribution of galaxies, support this possibility by indicating that
most galaxies at high $z$ have properties that would tend to repel
dust grains, out to a radius large enough that the low-density IGM can
be significantly polluted.  The chief uncertainties in our calculation
are {\em not} the detailed choices of dust opacity, IMF, dust
correction, cosmological parameters, etc. (all of which are probably
uncertain only at a level which does not significantly affect our
results), but rather in the physics of dust ejection itself.  We
assume that most dust reaches the point of equilibrium between
radiation pressure and graviational forces, but realistically dust
might be destroyed in transit, or confined to galaxies by other
forces.  Gas drag can confine grains at small galactic scale-heights,
but this still allows a large dust outflow when the circulation of gas
in the galaxy is considered (Shustov \& Vibe 1995).  But magnetic
fields (not included in our treatment) might be extremely important,
since charged grains in a microgauss field would oscillate about field
lines with a Larmor radius significantly smaller than the galaxy
scale.  To escape, dust may diffuse along a vertical component of the magnetic
field (Shustov \& Vibe 1995), perhaps enhanced by low-level winds or
by Parker instabilities (Chiao \& Wickramasinghe 1972; Ferrara et
al. 1991).  Magnetic fields may also be much weaker at high-z if they
have been amplified by a dynamo since then.

If dust {\em can} escape magnetic fields, our calculations show that
 it could significantly pollute the IGM.  A unique signature of
 enrichment by dust is that while dynamics or winds would pollute the
 IGM with chemical abundances similar to those of the galaxies, dust
 ejection can only enrich the IGM with elements such as C, Si, and Mg,
 which solidify as grains.  Elements such as N, Zn, and the noble
 gases, which are very lightly depleted onto grains, should only be
 ejected in trace amounts.  Thus by measuring the relative ratio of N
 to C or Si in Ly$\alpha$ lines, one could constrain the pollution by
 dust.  Presently N is detected only in absorbers of fairly high
 ($N(H\,I) \sim 10^{16}\,{\rm cm^{-2}}$) column density (Songaila \&
 Cowie 1996), but pushing these observations to lower H\,I columns
 could give strong constraints on (or evidence for) dust
 enrichment.\footnote{Unfortunately (for this application), N might
 also be lacking if it is underproduced in the massive (perhaps
 low-metallicity) stars responsible for the enrichment at high-$z$;
 see Arnett (1995).}  At low redshifts the significant abundances of
 Ne and Ar in cluster gas (e.g., Mushotsky et al. 1996) indicates that
 dust cannot be the sole pollutant of the ICM and that some enrichment
 by other mechanisms must occur.  Higher quality data from {\em
 Chandra} should allow a much more interesting test of the importance
 of dust ejection.

	Our calculations also give a fairly accurate assessment of the
expected opacity of rich clusters if about half of the observed enrichment
were due to dust ejection (unless the grain opacities are significantly higher than
we have assumed). Our estimate of $\sim 0.2-0.5$\,mag in the central few
hundred kpc of clusters is roughly comparable to both claimed detections
of cluster dust using extinction of background quasars (e.g., Boyle et
al. 1988; Romani \& Maoz 1992) or IR emission (Stickel et al. 1998),
and to upper limits based on reddening (e.g. Maoz 1995).\footnote{The
conclusions of reddening studies are vulnerable to changes in the dust
grain-size distribution by dust destruction; see Aguirre (1999).}  This
indicates that the general picture of substantial dust ejection from
galaxies might provide an interesting level of extinction through the
IGM, but would not violate any current constraints on cluster dust
density.

The primary difficulty with dust ejection as an explanation for the Si
and C enrichment of the low-density IGM is that it is not at all clear
-- theoretically or observationally -- that dust really can decouple
from galactic gas; but if it can, there appears to be no reason why it
would not escape to large radii.  Galactic winds, on the other hand,
are clearly observed both locally and at high-$z$, and they should
certainly be able to pollute (at least) the halos of their progenitor
galaxies.  But spreading the metals to large distances may disrupt the
IGM more than observations allow (e.g., Theuns, Mo \& Schaye 2000; Aguirre
et al. 2001b).  This suggests a possible `hybrid' scenario in which
gas and dust are expelled into a diffuse mixture in the halos of
galaxies.  But while gas remains there, the dust could continue,
driven by radiation pressure, to large distances.

	For example, imagine a representative $z=5$ galaxy of baryonic
mass $5\times 10^9\msol$ and UV-optical-NIR luminosity $2.5\times
10^{10}\,L_\odot$ driving a wind of velocity $v=200\kms$ at small
radii.  Such a galaxy could reasonably drive a wind to $\sim
1-100\,$kpc (using the results discussed in Aguirre et al. 2001b), but
our calculations show that radiation pressure could exceed
gravitational attraction out to 100-200\,kpc.  If $0.1\mic$ graphite
grains were to decouple from the gas near the disk (say at $10$\kpc),
they could reach $(100-200)\kpc$ after $\approx(0.2-0.4)$Gyr, with
velocity $\approx 470-500\kms$ (in calculating this we assume here
that the enclosed mass is proportional to the radius). After
$\sim1$\,Gyr (i.e. at $z=3$) the grains could reach up to
$\sim420\kpc$; silicate grains could reach up to $\sim280\,$kpc during
the same time. This is an upper limit since we have neglected gas
drag; adding gas drag appropriate for $\delta=100(10)$ gas
reduces the graphite distance to 250(385)\,kpc and the silicate
distance to 220(270)\,kpc.\footnote{For this estimate we use a drag force
$\sigma\rho v^2$ where $\sigma$ and $v$ are the dust geometrical cross
section and velocity, and $\rho$ is the density of the medium.  This
neglects Coulomb drag and is good in the limit of highly supersonic
grains. }  but indicates that radiation pressure can quite plausibly
eject dust far enough to pollute the IGM quite uniformly while
disturbing the IGM only near the galaxy.

In summary, our calculations indicate that galaxies at high redshift
tend to repel rather than attract dust grains.  If a substantial
fraction of dust can reach at least the equilibrium radius between
gravitational and radiation pressure forces, then the ensuing
enrichment can account for the mean level of C and Si observed in the
IGM at $z\sim3$.
Dust ejection would also enrich groups and clusters
substantially, though radiation pressure cannot account for {\em all}
of the metals observed.  The resulting dust extinction would be $\la
0.5$\,mag through the cores of rich clusters.  Dust ejection and the
ejection of metals by winds are, in some sense, complementary. Winds
almost certainly drive gas into the halos of galaxies, but may
overly-disturb the IGM if the gas travels to very large radii.  Dust
may be confined to galaxies by magnetic fields or gas drag but should
leave unimpeded if first moved into the halo.  Radiation pressure
acting on dust can therefore help enrich the IGM more uniformly than
winds alone.  Because only certain elements form dust, the possibility
of intergalactic enrichment by dust can be robustly tested -- in
principle -- by measuring ratios between refractory and non-refractory
elements in the IGM.

\acknowledgements

This work was supported by NASA Astrophysical Theory Grants NAG5-3922,
NAG5-3820, and NAG5-3111, by NASA Long-Term Space Astrophysics Grant
NAG5-3525, and by the NSF under grants ASC93-18185, ACI96-19019, and
AST-9802568.  JG was supported by NASA Grant NGT5-50078 for the
duration of this work, and AA was supported in part by the National
Science Foundation grant no.\ PHY-9507695 and by a grant in aid from
the W.M. Keck Foundation.  The simulations were
performed at the San Diego Supercomputer Center.

\end{document}